\documentclass[12pt]{article}
\usepackage{graphicx}
\usepackage{epsfig}
\begin{document}
\def\BibTeX{\rm B{\sc ib}\TeX}

%\preprint{Simon et al.}

\title{Superlattice of resonators on monolayer graphene created by intercalated gold nanoclusters}

\author{M. Cranney$^1$, F. Vonau$^1$, P.B. Pillai$^2$, E. Denys$^1$, D. Aubel$^1$, \\M.M. De Souza$^2$, C. Bena$^{3,4}$ and L.
Simon$^1$\footnote{corresponding author, Email address: L.Simon@uha.fr $^1$}
\\
{\small \it $^{1}$Institut de Sciences des Mat\'{e}riaux de
Mulhouse IS2M-LRC 7228-CNRS-UHA,}\vspace{-0.1in}
\\{\small \it 4, rue des fr\`eres Lumi\`ere, 68093 Mulhouse-France}
\\{\small \it $^2$Semiconductor Materials
and Device group Electronic} \vspace{-0.1in}\\{\small \it and Electrical Engineering University of Sheffield,}
\vspace{-0.1in}\\{\small \it Mappin street S1 3JD Sheffield,
UK} \\{\small \it$^3$Laboratoire de Physique des Solides, Universit\'{e} Paris-Sud 11,} \vspace{-0.1in}\\{\small \it 91405 Orsay Cedex, France} \\{\small \it $^4$Institut de Physique Th\'{e}orique, CEA/Saclay, CNRS-URA 2306,} \vspace{-0.1in}\\{\small \it Orme des Merisiers, 91191 Gif-sur-Yvette Cedex, France}}

\maketitle

\begin{abstract}
Here we report on a ``new'' type of ordering which allows to modify
the electronic structure of a graphene monolayer (ML). We have
intercalated small gold clusters between the top monolayer
graphene and the buffer layer of epitaxial graphene. We show that
these clusters perturb the quasiparticles on the ML graphene, acting as quantum
dots creating a superlattice of resonators on the graphene ML, as
revealed by a strong pattern of standing waves.
A detailed analysis of the standing wave pattern using Fourier
Transform Scanning Tunneling Spectroscopy strongly indicates that
this phenomenon can arise from a strong modification of the band
structure of graphene and (or) from Charge Density Waves (CDW)
where a large extension of Van Hove singularities are involved.

\end{abstract}

%\pacs{68.65.-k, 81.16.Fg, 81.07.-b, 81.16.Rf, 82.30.RS, 82.65.+r}

Graphene is a system exhibiting massless quasiparticles
which are able to propagate ballistically over mesoscale distances
\cite{NovoselovNature2005}. The dispersion of graphene quasiparticles can
to a first approximation be easily described by the Dirac
equation. The Dirac quasiparticles at low energy have a conic dispersion
around the K points. At high energy however, a band crossover occurs at the M
points of the Brillouin zone, such that the electron-like band dispersion becomes
hole-like beyond an energy equal to first approximation to the hopping energy;
this crossover gives rise to  a Van Hove singularity (VHS) in the tunneling density of states.
Due to this particular topology of the constant
energy contours (CEC) in the vicinity of the Van Hove singularity, graphene is a
good candidate to study the possibility of engineering a high temperature superconductor
via the so-called Van Hove scenario. Indeed, in this scenario which remains controversial
\cite{Bouvier2010}, a strong electron-phonon coupling is expected
with a large extension of the Van Hove singularities.

In the
context of harnessing and exploring these interesting topological Van Hove singularities,
a major challenge is to manipulate the graphene
layer while preserving its properties as much as
possible. Recently, the heavy doping of epitaxial monolayer
graphene has allowed to bring the Fermi level close to the energy of the Van Hove singularities;
subsequent ARPES measurements have shown strong trigonal
warping effects \cite{RotenbergPRL2010}. It has also been
demonstrated that the rotation between two stacked graphene layers can
tune the energy of the Van Hove singularities
\cite{Andreinaturephysics2010}. Another more general approach is
to create a periodic structure of quantum units that generates a new
``metamaterial''. Indeed, it has been demonstrated very recently the
possibility to create a new dispersive band structure with the
confinement of Shockley states on a copper surface using a
periodic network of supramolecules \cite{St?rhScience2009}. Such
an elegant method remains hypothetic for graphene because of the
behavior of the Dirac electrons and of the so-called Klein paradox
which predicts that potential barriers are transparent to
electrons, and sophisticated lithography techniques are the
only realistic method to generate such periodic structures.

Here
we report on a ``new'' type of ordering which allows a modification
of the graphene layer electronic structure. We have intercalated
small gold clusters between the top monolayer graphene and the
buffer layer of epitaxial graphene. We show that these clusters
perturb the quasiparticles of the ML graphene and act as quantum dots creating
a superlattice of resonators, as revealed by a
strong pattern of standing waves on the ML graphene. This is observed
in a specific range of positive energies where the Van Hove singularities
could be involved via a large Van Hove extension as found by
McChesney et. al. \cite{RotenbergPRL2010}. A deeper insight into the
standing-waves pattern using the Fourier Transform of dI/dV map Scanning Tunneling Microscopy (FT-STS) images  strongly suggests a substantial modification of the band structure of graphene and (or)
reveals the formation of charge density waves (CDW). The advantage of
investigating the VHS via our method versus ARPES is that ARPES
probes the occupied states only, thus the study of positive energy VHS needs high
levels of doping which can affect the band dispersion of graphene
via hybridization of its bands with the dopant atoms
\cite{GuineaPRB2008}. In our case the FT-STS allows to probe also
the empty states {\it without} the possible modifications induced by the dopants on the
dispersion band of graphene. Our ability to fabricate large
homogeneous structures with a high amount of control and reproducibility,
as well as the observation of the quantum confinement,
which has not been observed before in graphene, open the path
for multiple theoretical and experimental explorations. Moreover, since
graphite intercalation compounds display superconducting
properties, our new metal intercalation method can have applications
to high Tc CDW materials.

\begin{figure}
\includegraphics[width=5in]{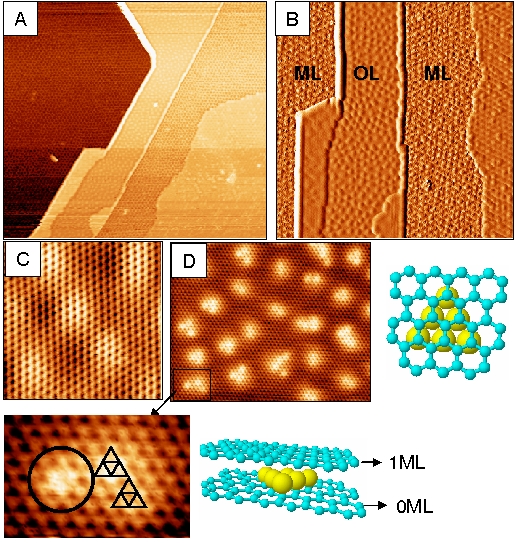}\\
\caption{ STM picture at 77 K of the surface of epitaxial graphene
obtained after 1 ML gold deposition followed by 5 min of annealing
at 1000 K. A) Large area topographic image ($138\times124$
$nm^{2}$, V=-1.42V) showing the frontier between two steps and the
modification of the ML graphene due to the intercalation of gold.
In B) a zoomed image of A) ($67\times65$   $nm^{2}$, -1.5V) shows
the pristine monolayer graphene (ML) and the ``ostrich leather''
(OL) region which consists in the intercalation of aggregates of
flat $Au_{6}$ clusters under the first monolayer graphene as
zoomed in D) and schematized on the nearby figures. C) Zoomed
topographic image of OL region at low bias voltage ($5\times5$
$nm^{2}$, -110 meV) showing the six atoms of the honeycomb
structure which ascertains that aggregates are inserted under a
monolayer graphene. (image processing using WSxM software
\cite{wsxm})} \label{Fig1}
\end{figure}

The graphene samples were prepared in UHV by the annealing of
n-doped SiC(0001) at 900 K for several hours and subsequent
annealing at 1500 K \cite{VanBommelSurfSci75,
ForbeauxPRB98,SimonPRB99,others}. This preparation method leads to
the formation of a buffer graphene layer covalently bonded with
the substrate and a ML graphene decoupled from the substrate
\cite{LaufferPRB08}. The epitaxial graphene has an intrinsic
n-type character and the Dirac point is at 0.4 eV below the Fermi
level \cite{OhtaSci06, BergerSci06}. In a recent report, a simple
way to shift the Fermi level to the unoccupied states of epitaxial
graphene and induce p-type doping by deposition of metal atoms on
top of graphene was proposed \cite{GierzNanolett08}. Contrary to
the admitted idea that the metallic atoms stay on the surface and
are covalently bonded with the carbon atoms, we have shown that
gold atoms do not stay on the graphene surface but intercalate
between the buffer layer (silicon carbide substrate plus the 0ML
graphene) and the first true ML graphene \cite{PremlalAPL09}.
Depending on the analyzed region, we find a doping effect
associated to the formation of a complete gold monolayer
\cite{forcomingpaper} and another region called ``Ostrich Leather'' which is a pseudoperiodic repartition of aggregates of gold
nanoclusters as displayed in Figs \ref{Fig1} A) and B). In the
latter case no doping effect was found. Here we will focus on this
particular clustering phase which creates the standing wave
pattern. The deposition of gold on graphene was carried out at
room temperature using a homemade Knudsen cell calibrated using a
Quartz Crystal Microbalance. The sample was further annealed at
1000 K for 5 min. Our experiments were performed with a LT-STM
from Omicron at 77 K at a base pressure in the $10^{-11}$ mbar
range. Every image was acquired using a lock-in amplifier and a
modulation voltage of $\pm 20 mV$.

At low bias voltage the atomic resolution reveals the six atoms of
the honeycomb structure (see figure \ref{Fig1} C)) which
ascertains that the top layer is ML graphene. At higher negative
bias voltages, the aggregates are more visible and the image
reveals that each protrusion consists of aggregates of smaller
clusters of gold atoms (figure \ref{Fig1}D)). The measured height
difference between pure ML graphene and the OL layer is $0.8\pm
0.1\AA$. This corresponds to the height difference between a ML
layer and a bilayer graphene formed on two adjacent steps with the
bilayer graphene formed on the lower terrace \cite{LaufferPRB08}.
This signifies that the space between the top ML graphene and the
buffer layer does not allow the insertion of 3D gold islands.
Therefore we propose that the Au aggregates consist of small flat
clusters. A zoomed image of figure \ref{Fig1} D) indicates
triangular shaped clusters with a size corresponding to 6 gold
atoms as already found in the equilibrium gas phase
\cite{Hakkinen08} and schematized figure \ref{Fig1}. The cluster
structure is only visible at higher negative bias voltage and does
not show any specificity in dI/dV map.

\begin{figure}
\includegraphics[width=5in]{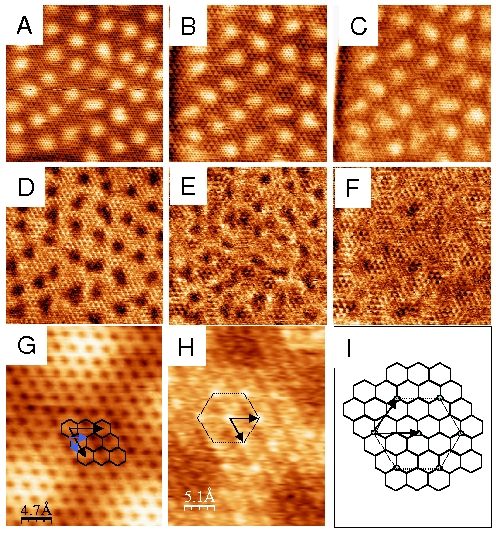}\\
\caption{ Topographic A)-C) and dI/dV maps D)-F) of a OL region
for the bias voltage from +0.8 to +1.0 V (empty states) which
develops a full standing wave pattern ; A), D) ($14\times14$
$nm^{2}$) and B), C), E) and F) ($12.5\times12.5$ $nm^{2}$). G)
and H) display a zoom of a resonators region which shows that
standing waves manifest as bright protrusion with a p(2x2)
reconstruction as schematized in I). }\label{Fig2}
\end{figure}

\begin{figure}
\includegraphics[width=5in]{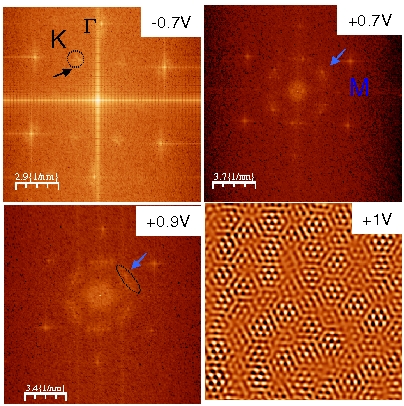}\\
\caption{ 2D Fourier Transform (power spectrum) of the dI/dV map
image at the energy indicated on the figures. Black and blue
arrows indicate respectively the K and M positions in the first
Brillouin Zone. The image at +1.0V corresponds to the FFT
filtering of the dI/dV image taking only the elliptical feature
around M point has indicated on the FT-STS image at +0.9V. This
gives a filtered dI/dV image where the observed network of
resonators is strongly enhanced. } \label{Fig3}
\end{figure}

Figure \ref{Fig2} shows a region of OL for positive bias voltage
at +0.8, +0.9 and +1.0 V respectively. Here we probe the empty
states. From A) to C) the topographic images show the clusters as
large protrusions and from D) to F) strong standing wave pattern
are observed. It seems that the aggregates create an excess of
electrons and appear as dark regions for the empty states energies
in the dI/dV maps. The size of the dark regions is reduced with an
increase in bias energy and finally a complete strong and
contrasted standing wave pattern is observed in F). From +0.8 eV
to 1 eV the size changes linearly with the energy. The zoomed
topographic image G) and dI/dV map H) show that the standing wave
manifest as bright protrusions with lattice vectors which
correspond to a p-2x2 (in the same directions as the unit cell
vectors as schematized in I). Figure \ref{Fig3} displays the 2D
power spectrum of the 2D FT of the dI/dV map images at bias
voltages of -0.7V, +0.7V, +0.9V. As first shown by Sprunger et al
\cite{SprungerScience97}, considering the FT of the dI/dV map of
the standing wave pattern, the 2D power spectrum gives a direct
representation of the constant energy contour (CEC) at a given
energy. It provides a direct representation of wavelengths of
standing waves in the local density of states and also details of
the possible scattering process. We have demonstrated that the
power spectrum of the 2D FT-STS is in a first approximation the
Joint Density of States (JDOS) at a given energy which can be rapidly
calculated by the 2D self-correlation function of the CEC
\cite{JofCondMater07}. The uncertainty is given by the bias
modulation of the lock-in acquisition (here $\pm20meV$). Provided
that bands are sufficiently robust, we have demonstrated that this
technique allows a complete determination of the energy dispersion of
a 2D semimetal by the interpretation of the features shapes and
their modifications with energy \cite{VonauPRL05}. This approach
has been successfully used on superconductors
\cite{McElroyNature03}, and more recently on epitaxial graphene in
order to measure the group velocity of the QPs
\cite{RutterScience07} and to provide a direct evidence of the
chiral property of electrons in monolayer graphene
\cite{MalletPRL08}. Since two waves with different spin
orientations can not interfere, the absence of expected features
associated with specific orientation and length of momentum
indicates that the joined points of the CEC have different spins
\cite{PascualPRL04, RoushanNature09}. 

At -0.7V faint features
around K points appear. For the energy of +0.7V, a stronger
feature appears around M as indicated in figure 3. This
corresponds to a manifestation of the standing wave pattern as
observed and discussed in figure \ref{Fig2}. Indeed, the FFT
filtering of the dI/dV map in figure \ref{Fig3} at +1.0V, taking
into account only the elliptical feature around M points
(indicated by the blue arrow on the neighboring at +0.9V), gives a
filtered dI/dV image where the observed network of resonators is
strongly enhanced. The features around M points in FT-STS become
more and more contrasted and elliptical with the long axis of the
ellipse along the direction $M-K-\Gamma$ and further develop into
distinct bright spots (two at +0.7V and four at +0.9V). Due to the
intrinsic n-type doping if we expect a large extension of the
Van Hove singularities as reported in \cite{RotenbergPRL2010}, the
origin of these features could be compatible with the idea of
localization of the density of states around M points due to a
band crossover between two consecutive K and K' contours as shown
in figure \ref{Fig4}. In A) we represent the calculated JDOS image
which corresponds to the CEC given in D) below the crossover
energy. At these energies the contours are localized around the K
points. JDOS shows features around the K points as currently
expected and identical to the ones observed in figure \ref{Fig3}
at the negative energies. In B) the JDOS pattern is calculated for
the theoretical CEC of a graphene monolayer represented in E) at
the energy which corresponds to the band crossover when K contours
touch each other as schematized in F). This energy corresponds to
the so-called Van Hove singularities. In C) a zoom of the
calculated JDOS around M point shows two features. An elliptically
shaped broad feature with the long axis along $M-K-\Gamma$
direction and two more contrasted points in the perpendicular
direction i.e. along the $\Gamma-M-\Gamma$ direction are observed. As shown by
the two blue arrows the elliptical feature can only be associated
with the momentum which joins two equivalent points of two
consecutive CECs around the K points (intervalley scattering). The
feature along the $\Gamma-M-\Gamma$ direction (red arrow) as
obtained in the calculated JDOS in B) should be associated with
the momentum which joins two corners of the same triangular shaped
K contour as shown in E) (intravalley scattering). The
experimental FT-STS (see figure 3 at +0.9V) however displays only
the ellipse and the spots predicted by theory and indicated by red
arrows are not observed.

To explain the FT-STS features, we have used a simple theory
based on JDOS calculations, for which the band
structure was computed by considering only the first nearest-neighbor tight-binding
(1st order NN TB) hopping. Other factors such as higher order hopping terms, as well as the difference between a full T-matrix calculation and a JDOS calculation of the FT-STS spectra have also been considered but
since the qualitative aspects of the results are essentially unchanged we do not discuss these more complicated scenarios here.

We show that, in the JDOS approach, the elliptical features at the M points appear
in the spectra only if the CECs touch together (see the supplementary information).
We have tested two scenarios. For normally dispersing graphene, when the energy is decreased
below the Van Hove energy, the neighboring triangular CECs centered at the K points no longer intersect, and the ellipses at the M points cease to be observed as soon as we decrease the energy below the Van Hove energy. They are replaced by contours centered around the K point that diminish in size with decreasing energy. In the second scenario the Van Hove singularity is extended to a flat band, for which we ``nest''
the momentum at the Van Hove energy. In this situation the increase of energy
yields an increasing of the area of the flat band
in the vertex of the triangular CECs (as illustrated in fig.
\ref{Fig4} F) contributing to the DOS. Our simulations clearly show that the size of the
elliptical feature in the JDOS depends on the size of the black
area in the vertex of the triangular CEC (see the supplementary
information). This implies that the elliptical feature can be a
fingerprint of the extension of the Van Hove singularity.
Along the same lines, in the work of Rotenberg et. al.
\cite{RotenbergPRL2010}, where the possible link between doped
graphene and superconductivity is strongly supported, the ARPES measurements show that, as soon as the vicinity of the Van Hove
singularity is reached, the vertices of two consecutive triangular
CEC contours begin to touch, and the density of states
in the vertices is being ``filled'' by the doping (see fig. 1 G) and H) in \cite{RotenbergPRL2010}).
The situation appears quite similar to our observation depicted in
figure \ref{Fig4} F) which corresponds also to a non-conventional
dispersion, however, in our situation for a non highly doped graphene.

The elliptical features at the M points seem to be non-dispersive,
i.e. their position does not change in the energy range from +0.6
to +1 eV. This p(2x2) standing wave pattern could thus also have a
CDW character. Indeed, this situation is also similar to the
observation by STM of the so-called ``checkerboard'' structures on
high Tc superconductors which reveal an ordering structure of four
unit-cell wavelength \cite{McElroyNature03}. Note that quasiparticle
interference (QPI) processes associated to spatial LDOS modulation
(Friedel oscillations) can in general be distinguished from CDWs
as the latter are non dispersive.

Our results raise many theoretical and experimental questions. The
first question concerns the physical significance of the contrasted
point-like structures observed in the elliptical region in the
FT-STS. Such intensity enhancement can also be found in the ARPES
measurements \cite{RotenbergPRL2010}. Our calculations of the
FT-STS feature indicate that this may be the result of some extra pockets
of states that could be created along the M-K direction;
more theoretical studies are needed however to explain the origin of such unusual features.
The second question concerns the
absence of the strong features expected in the JDOS calculation
along the direction perpendicular to the observed ellipses (red
arrows in figure 4 C) and E)). Similar to the nodal-antinodal
dichotomy expected for the hole doped high Tc superconductors, the quasiparticles at the Van Hove singularities are well
defined along the nodal direction (in our case $\Gamma-M-\Gamma$ direction),
while in the anti-nodal direction they are broad, even
incoherent (see figure \ref{Fig4} C)). The similarity of the two phenomena is striking. A comparison of our
experimental observation with the JDOS scheme of the expected scattering events reveals that the nodal-direction events are visible (blue arrows),
while those lying in the antinodal direction are not (red arrows). Thus the JDOS approach appears to
reproduce some of the experimental features, but fails to fully explain all the FT-STS features,
particularly the presence of the bright spots inside the
elliptical features around M, as well as the absence of intensity along the perpendicular directions. Such features are not recovered by more complex approaches such as the T-matrix, and more investigation is necessary to identify their possible origins.

The last, but not the least outstanding question concerns the ability of the gold intercalated
clusters to act as efficient scatterers. Usually monolayer
graphene does not exhibit such contrasted standing wave patterns,
except in the presence of strong point defects. Here it is assumed that the gold clusters do not act as point-like
scatterers, but they screen the charge transfer from the silicon carbide
substrate to the monolayer graphene. This should not create an
effective doping of the graphene layer since the gold layer is
discontinuous \cite{PremlalAPL09}, but each cluster should act as a
quantum dot creating a pseudo graphene superlattice similar to the nanoporous quantum network obtained via the adsorption
of molecules on copper surface in Ref.~\cite{St?rhScience2009}.

To conclude, our results open the path for further theoretical and experimental explorations. More experimental studies such as ARPES and transport are required to clarify the nature
of the standing-wave patterns observed in graphene in the presence of the superlattice of gold islands. Moreover, from a theoretical perspective other studies such as ab-initio, and LDA may help to identify the changes induced by the intercalated gold atoms to the band structure of graphene, and help one to distinguish between the scenarios of a CDW-like ordering of the quasiparticles, and Friedel oscillations in an extended Van Hove singularities scenario. It would also be interesting to study the possibility of a connection to the physics of high temperature superconductors,
as well as a possible tailoring of the Van Hove singularities by the ``activation'' of
standing waves. Moreover, from a more applied perspective,
the superlattices created here could be used as a trap for impurities or
molecules in order to improve the self-organization process. Furthermore, they open the way towards a controlled molecular nanostructuration using graphene layers. Last but not least,
it would be interesting to generalize the type of intercalation presented here for other metallic elements.

\begin{figure}
\includegraphics[width=5in]{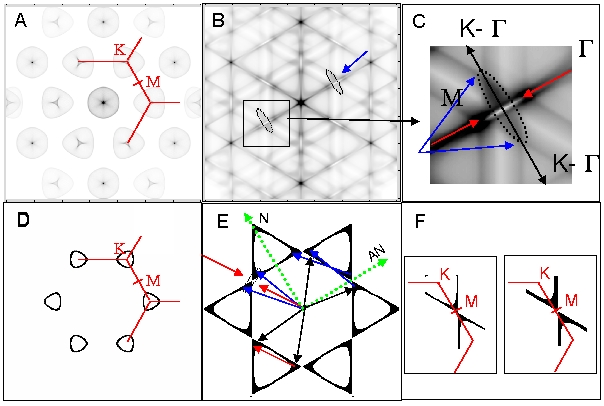}\\

\caption{A) and B) Calculated 2D Fourier Transform (power
spectrum) for the theoretical CECs contours given in D) and E)
respectively in the case of pristine monolayer graphene (1NN
tight-binding model). A zoom of the calculated features found
around M point is given in C). In E) blue and red arrows
correspond to the momentum vectors associated to the black
vectors. In C) these vectors are reported in the zoomed features
around the M points : the elliptical feature by blue arrows and
the two structures in the perpendicular direction by red arrows
around the M point in the calculated JDOS. This gives a direct
interpretation of the features found in the calculated JDOS in B).
Blue arrows which describe the ellipse correspond to intervalley
scattering between two first nearest neighbor CEC contours. Red
arrows (not observed experimentally) would correspond to
intravalley scattering process (see in E)).  F) gives the proposed
CEC evolution with increasing energy in the vicinity of the
Van Hove singularity explaining the increase of the elliptic size
found around M in B) and C). The size of the ellipse increases
with the area of the density of states in the vertex of the
triangular shaped CEC contours.} \label{Fig4}
\end{figure}

Acknowledgement: This work as part of the European Science
Foundation EUROCORES Programme FoNE, was supported by funds for
exchange visit by Mr. P.B. Pillai to Mulhouse, from the EPSRC and
the EC Sixth Framework Programme, under Contract N.
ERAS-CT-2003-980409. CB would like to acknowledge funding from the
French ANR (Agence Nationale de Recherche), under the P'NANO
program, reference NANOSIM-GRAPHENE.

\begin{figure}
\includegraphics[width=6.5in]{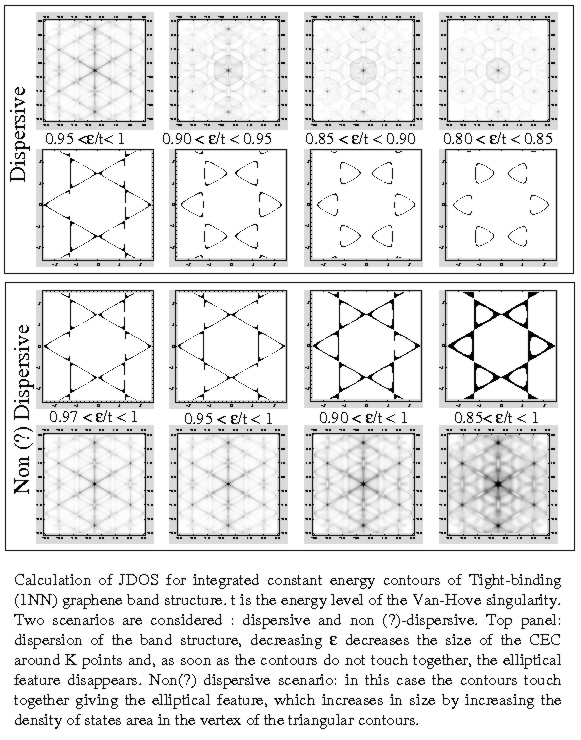}\\
%\vspace{-1in}
\caption{Supplementary information} \label{Fig4}
\end{figure}

\end{document}